# Investigation of $g_{f_0\rho\gamma}$ and $g_{a_0\rho\gamma}$ coupling constants in light cone sum rules

C Aydin[*], M Bayar and A H Yilmaz[†]

*Physics Department, Karadeniz Technical University, 61080 Trabzon, Turkey*
[*] coskun@ktu.edu.tr
[†] hakany@ktu.edu.tr

**Abstract**

We present a calculation of the coupling constant of $f_0 \to \rho\gamma$ and $a_0 \to \rho\gamma$ decays from the point of the light cone QCD sum rules. We estimate the coupling constants $g_{f_0\rho\gamma}$ and $g_{a_0\rho\gamma}$ which are an essential ingredient in the analysis of physical processes involving isoscalar $f_0(980)$ and isovector $a_0(980)$ mesons.

Keywords: coupling constant, isoscalar $f_0(980)$, isovector $a_0(980)$, $\rho(980)$, light cone QCD sum rule

**I. Introduction**

To determine the fundamental parameters of hadrons from experiments, some information about physics at large distances is required. As known that the large distance physics cannot be calculated directly from the fundamental QCD Lagrangian because at large distance perturbation theory cannot be applied. For this reason a reliable nonperturbative approach is needed. The QCD sum rules [1] have proved to be very useful to extract the low-lying hadron masses and coupling constants. This method is a framework which connects hadronic parameters with QCD parameters. It is based on short-distance operator product expansion (OPE) in the deep Euclidean region of vacuum–vacuum correlation function in terms of quark and gluon condensates. Further progress has been achieved by an alternative method known as the QCD sum rules on the light cone [2-5]. The light-cone QCD sum rules method has also been used to analyze several hadronic properties and to calculate hadronic coupling constants. The method of light-cone sum rules is a hybrid of the standard technique of QCD sum rules in the manner of Shifman, Vainshtein, and Zakharov (SVZ) [1] with the conventional distribution amplitude description of the hard exclusive process [6,7]. The basic idea of SVZ sum rules is to use vacuum condensates to parametrize the nontrivial QCD vacuum and employ the duality hypothesis to relate the experimental observables to the theoretical calculation. Technically, operator product expansion (OPE) based on the canonical dimension is used. The difference between SVZ sum rules and light-cone sum rules is that the short-distance Wilson OPE in increasing dimension is replaced by the light-cone expansion in terms of distribution amplitudes of increasing twist.

With increasing experimental information about the different members of the meson spectrum it becomes very important to develop a consistent understanding of the observed mesons from a theoretical point of view. For the low-lying pseudoscalar, vector, and tensor mesons this has been done quite successfully within the framework of the simple quark model assuming the mesons to be quark-antiquark $(q\bar{q})$ states grouped together into nonets. For the scalar mesons, however, several questions stil remain to be answered, most of them related to the nature of the experimentally observed mesons $f_0(980)$ and $a_0(980)$.

The light mesons have been the subject of continious interest in hadron spectroscopy. Although the structure of these light mesons have not been unambiguously determined yet [8-10], the possibility may be suggested that these nine scalar mesons below 1 GeV may form a scalar SU(3) flavor nonet [11]. In the naive quark model, we have $a_0 = (u\bar{u} - d\bar{d})/\sqrt{2}$ and $f_0 = s\bar{s}$, while in the framework of four-quark models, the mesons $f_0(980)$ and $a_0(980)$ could either be compact objects i.e. nucleon-like bound states of quarks with the symbolic quark structures $f_0 = s\bar{s}(u\bar{u} + d\bar{d})/\sqrt{2}$ and $a_0 = s\bar{s}(u\bar{u} - d\bar{d})/\sqrt{2}$ [12-15] or spatially extended objects i.e. deuteron-like bound states of hadrons; the $f_0(980)$ and $a_0(980)$ mesons are usually taken as *KK* molecules [9, 16, 17]. The nature of the meson $f_0(980)$ is particularly debated. One of the oldest suggestions, there is the proposal that quark confinement could be explained through the existence of a state with vacuum quantum numbers and mass close to the proton mass [18]. While considering the strong coupling to kaons, $f_0(980)$ could be interpreted as an $s\bar{s}$ state [19-23]. On the other hand, it does not explain the mass degeneracy between $f_0(980)$ and $a_0(980)$ interpreted as a $(u\bar{u} \pm d\bar{d})/\sqrt{2}$ state. A four quark $qq\bar{q}\bar{q}$ state definition has also been offered [12,13]. The identification of the $f_0$ and of the other lightest scalar mesons with the Higgs nonet of a hidden $U(3)$ symmetry has also been proposed [24]. On the other hand, they are relevant hadronic degrees of freedom, and therefore the role they play in hadronic processes should also be studied besides the questions of theire nature. In this work, we calculated the coupling constant $g_{f_0\rho\gamma}$ and $g_{a_0\rho\gamma}$ by applying the light-cone sum rule, which provide an efficient and model-independent method to study many hadronic observables, such as decay constants and form factors in terms of non-perturbative contributions proportional to the quark and gluon condensates [5]. On the other hand we are also calculating the same coupling constants in the framework of QCD sum rules.

Radiative transitions between pseudoscalar (P) and vector (V) mesons have been an important subject in low-energy hadron physics for more than three decades. These transitions have been regarded as phenomenological quark models [25], potential models [26], bag models [27], and effective Lagrangian methods [28, 29]. Among the characteristics of the electromagnetic interaction processes $g_{VP\gamma}$ coupling constant plays one of the most important roles, since they determine the strength of the hadron interactions. Low-energy hadron interactions are governed by nonperturbative QCD so that it is very difficult to get the numerical values of the coupling constants from the first principles. For that reason a semiphenomenological method of QCD sum rules can be used, which nowadays is the standart tool for studying of various characteristics of hadron interactions. On the other hand, vector meson-pseudoscalar meson-photon VPγ–vertex also plays a role in photoproduction reactions of vector mesons on nucleons. It should be notable that in these decays (V→Pγ) the four-momentum of the pseudoscalar meson P is time-like, $P'^2 > 0$, whereas in the pseudoscalar exchange amplitude contributing to the photoproduction of vector mesons it is space-like $P'^2 < 0$. Therefore, it is of interest to study the effective coupling constant $g_{VP\gamma}$ from another point of view as well.

## II. The Light-cone Sum Rules Calculation :

We choose the interpolating current for the $\rho$ and $f_0$ ($a_0$) mesons as

$$j_\mu^{\rho_0} = \frac{1}{2}\left(\bar{u}\gamma_\mu^a u^a + \bar{d}\gamma_\mu^a d^a\right) \quad , \quad J_{f_0} = \frac{1}{\sqrt{2}}\left(\bar{u}^b u^b + \bar{d}^b d^b\right)\sin\theta + \cos\theta\, \bar{s}s \quad , \quad \text{and}$$

$J_{a_0} = \frac{1}{2}\left(\bar{u}^b u^b - \bar{d}^b d^b\right)$ [30] respectively. $\rho$-meson consist of $u$ and $d$-quarks, and $s$-quark does not contribute in this calculation. $J_\nu^\gamma = e_u(\bar{u}\gamma_\nu u) + e_d(\bar{d}\gamma_\nu d)$ is the electromagnetic current with $e_u$ and $e_d$ being the quark charges.

In order to derive the light cone QCD sum rule for the coupling constants $g_{f_0\rho\gamma}$, we consider the following two point correlation function

$$T_\mu(p,p') = i\int d^4x\, e^{ip'\cdot x}\langle 0|T\{j_\mu^\rho(x) j_s(0)\}|0\rangle_\gamma, \tag{1}$$

where $\gamma$ denotes the external electromagnetic field, and $j_\mu^\rho$ and $j_s$ and are the interpolating current for the $\rho$ meson and $f_0(a_0)$, respectively.

We therefore sature the dispersion relation satisfied by the vertex function $T_\mu$ by these lowest lying meson states in the vector and the scalar channels, and in this way we obtain for the physical part at the phenomenological level the Eq. (1) can be expressed as

$$T_\mu(p,p') = \frac{\langle 0|j_\nu^\rho|\rho\rangle\langle\rho(p')|S(p)\rangle_\gamma\langle S|j_s|0\rangle}{(p'^2 - m_\rho^2)(p^2 - m_s^2)}. \tag{2}$$

In this calculation the full light quark propagator with both perturbative and nonperturbative contribution is used, and it is given as [31]

$i\Im(x,0) = \langle 0|T\{\bar{q}(x)q(0)\}|0\rangle$

$$= i\frac{\slashed{x}}{2\pi^2 x^4} - \frac{\langle\bar{q}q\rangle}{12} - \frac{x^2}{192}m_0^2\langle\bar{q}q\rangle - ig_s\frac{1}{16\pi^2}\int_0^1 du\left\{\frac{\slashed{x}}{x^2}\sigma_{\mu\nu}G^{\mu\nu}(ux) - 4iu\frac{x_\mu}{x^2}G^{\mu\nu}(ux)\gamma_\nu\right\} + ...$$
$$\tag{3}$$

where the terms proportional to light quark mass $m_u$ or $m_d$ are neglected. After a straightforward computation we have

$$T_\mu(p,q) = 2i \int d^4 x e^{ipx} A(x_\rho g_{\mu\tau} - x_\tau g_{\mu\rho})\langle\gamma(q)|\bar{q}(x)\sigma_{\tau\rho}q(0)|0\rangle \tag{4}$$

where $A = \dfrac{i}{2\pi^2 x^4}$, and higher twist corrections are neglected since they are known to make a small contribution [4]. In order to evaluate the two point correlation function further, we need the matrix elements $\langle\gamma(q)|\bar{q}(x)\sigma_{\tau\rho}q(0)|0\rangle$. This matrix element can be expanded in the light cone photon wave function [32, 33]

$$\langle\gamma(q)|\bar{q}\sigma_{\alpha\beta}q|0\rangle = ie_q <\bar{q}q> \int_0^1 du e^{iuqx}$$
$$\times \{(\varepsilon_\alpha q_\beta - \varepsilon_\beta q_\alpha)[\chi\varphi(u) + x^2[g_1(u) - g_2(u)]] + [q.x(\varepsilon_\alpha x_\beta - \varepsilon_\beta x_\alpha) + \varepsilon.x(x_\alpha q_\beta - x_\beta q_\alpha)]g_2(u)\} \tag{5}$$

Where $e_q$ is the corresponding quark charge, $\chi$ is the magnetic susceptibility, $\varphi(u)$ is leading twist two and $g_1(u)$ and $g_2(u)$ are the twist four photon wave functions. The main difference between the tradiational QCD sum rules and light cone QCD sum rule is the appearance of these wave function. Light cone QCD sum rules corresponds to summation of an infinite set of terms in the expansion of this matrix element on the tradiational sum rules. The price one pays for this is the appearance of a priori unknown photon wave functions. After evaluating the Fourier transform for the $M_1$ structure and then performing the double Borel transformation with respect to the variables $Q_1^2 = -p^2$ and $Q_2^2 = -p'^2$, we finally obtain the following sum rule for the coupling constant $g_{f_0\rho\gamma}$

$$g_{f_0\rho\gamma} = \frac{1}{\sqrt{2}} \frac{m_\rho <\bar{u}u>}{\lambda_{f_0}\lambda_\rho} e^{m_\rho^2/M_1^2} e^{m_{f_0}^2/M_2^2} \{-M^2\chi\phi(u_0)E_0(s_0/M^2) + 4g_1(u_0)\}\sin\theta \tag{6}$$

and $g_{a_0\rho\gamma}$

$$g_{a_0\rho\gamma} = \frac{1}{6} \frac{m_\rho <\bar{u}u>}{\lambda_{a_0}\lambda_\rho} e^{m_\rho^2/M_1^2} e^{m_{a_0}^2/M_2^2} \{-M^2\chi\phi(u_0)E_0(s_0/M^2) + 4g_1(u_0)\} \tag{7}$$

where the function

$$E_0(s_0/M^2) = 1 - e^{-s_0/M^2} \tag{8}$$

is the factor used to subtract the continuum, $s_0$ being the continuum threshold, and

$$u_0 = \frac{M_2^2}{M_1^2 + M_2^2}, \quad M^2 = \frac{M_1^2 M_2^2}{M_1^2 + M_2^2} \tag{9}$$

with $M_1^2$ and $M_2^2$ are the Borel parameters in the $\rho$ and $f_0(a_0)$ channels.

## III. Numerical Calculation

In our calculations we used the numerical values $<\bar{u}u> = -0.014$ GeV$^3$, $m_{f_0} = 0.98$ GeV, $m_{a_0} = 0.98$ GeV, $\lambda_{f_0} = 0.18 \pm 0.02$ GeV$^2$ [34], $\lambda_{a_0} = 0.21 \pm 0.05$ GeV$^2$, [35], $m_\rho = 0.77$ GeV. We note that neglecting the electron mass the $e^+e^-$ decay width of $\omega$ meson is given as $\Gamma(\rho \to e^+e^-) = \frac{4\pi\alpha^2}{3}\left(\frac{\lambda_\rho}{3}\right)^2$. Then using the value from the experimental leptonic decay width of $\Gamma(\rho \to e^+e^-) = 7.02 \pm 0.11$ keV for $\rho$ [34,36], we obtain the value $\lambda_\rho = 0.17 \pm 0.03$ GeV$^2$ for the overlap amplitude $\rho$ meson. In order to examine the dependence of $g_{f_0\rho\gamma}$ and $g_{a_0\rho\gamma}$ on the Borel masses $M_1^2$ and $M_2^2$, we choose $M_1^2 = M_2^2 = M^2$. Since the Borel mass $M^2$ is an auxiliary parameter and the physical quantitites should not depend on it, one must look for the region where $g_{f_0\rho\gamma}$ ($g_{a_0\rho\gamma}$) is practically independent of $M^2$. Firstly, we determined that this condition is satisfied in the interval $1.4 \text{ GeV}^2 \leq M^2 \leq 2.4 \text{ GeV}^2$ for $g_{f_0\rho\gamma}$. The variation of the coupling constant $g_{f_0\rho\gamma}$ as a function of Borel parameter $M^2$ at different $\theta$'s and constant value of the continuum threshold as $s_0 = 2.0$ are shown in figure 1. Examination of this figure shows that the sum rule is rather stable with these reasonable variations of $M^2$. In the $f_0 \to \rho\gamma$ decay, we then choose the middle value $M^2 = 1.9 \text{ GeV}^2$ for the Borel parameter in its interval of variation and obtain the coupling constant of $g_{f_0\rho\gamma}$ for various $\theta$ angles as between $g_{f_0\rho\gamma} = 1.99 \pm 0.56$ and $g_{f_0\rho\gamma} = 3.86 \pm 0.97$, where only the error arising from the numerical analysis of the sum rule is considered.

We also use the numerical values mentioned above for the magnetic susceptibility $\chi = 3.15 \text{ GeV}^{-2}$ [37]. Using the conformal invariance of QCD up to one loop order, the photon wave functions can be expanded in terms of Gegenbauer polynomials; each term corresponding to contributions from operators of various conformal spin. Due to conformal invariance of QCD up to one loop, each term in this expansion is renormalized separately and the form of these wave functions at a sufficiently high scale is well known. In the previous works, twist-4 photon wave functions were used [32, 33]. Since they are not correct one should use new functions that are calculated by Ball, Braun and Kivel [37] and hence we have used the asymptotic forms of the photon wave function given by [37]

$$\varphi(u) = 6u(1-u),$$

$$g_1(u) = -\frac{a}{16} - \frac{h}{8} \qquad (10)$$

where $a = 40u^2(1-u)^2$ and $h = -10(1 - 6u + 6u^2)$ [38]. We will set $M_1^2 = M_2^2 = 2M^2$ which sets $u = u_0 = 1/2$. Note that in this approximation, we only need the value of the wave functions at a single point; namely at $u_0 = 1/2$ and hence the functional forms of the photon wave functions are not relevant.

In Fig. 2, we also showed the dependence of the coupling constant $g_{f_0\omega\gamma}$ on parameter $M^2$ at some different values of the continuum threshold as $s_0 = 1.6$, 1.8 and 2.0 GeV$^2$ at $\theta = 30°$. Since the Borel masses $M_1^2$ and $M_2^2$ are the auxiliary parameters and the physical quantities should not depend on them, one must look for the region where $g_{f_0\omega\gamma}$ is practically independent of $M_1^2$ and $M_2^2$. We determined that this condition is satisfied in the interval $1.4 \text{ GeV}^2 \leq M_1^2 \leq 2.4 \text{ GeV}^2$. Choosing the middle value $M^2 = 1.9 \text{ GeV}^2$ for the Borel parameter in this interval of variation and we have the coupling constant of $g_{f_0\rho\gamma}$ for different $s_0$ values as between $g_{f_0\rho\gamma} = 1.97 \pm 0.57$ and $g_{f_0\rho\gamma} = 1.87 \pm 0.54$ in the calculation of light-cone sum rules. In Fig 3 it is given the dependence of the coupling constant $g_{a_0\rho\gamma}$ on parameter $M^2$ at some different values of the continuum threshold as $s_0 = 1.6$, 1.8 and 2.0 GeV$^2$. We have the interval of $1.4 \text{ GeV}^2 \leq M_1^2 \leq 2.0 \text{ GeV}^2$ then choosing the middle value $M^2 = 1.7 \text{ GeV}^2$ for the Borel parameter in this interval of variation and we have the coupling constant of $g_{a_0\rho\gamma}$ for different $s_0$ values as between $g_{a_0\rho\gamma} = 0.82 \pm 0.34$ and $g_{a_0\rho\gamma} = 0.85 \pm 0.36$. The coupling constant $g_{a_0\rho\gamma}$ was calculated [34] as $2.0 \pm 0.50$ and $1.30 \pm 0.30$ in QCD sum rules.

It follows from Figures 2 and 3 that for the choises between $s_0 = 1.6$, 1.8 and 2.0 GeV$^2$ the variations in the result is about ~10%, ie., the coupling constant can be said to be essentially independent of the continuum threshold $s_0$. Furthermore, the coupling constant seems to be intensitive to variation of the Borel parameter $M^2$. Where all possible sourses of uncertainties are taken into account, namely, errors coming from determination of $\lambda_{f_0}$, from variation of the continuum threshold $s_0$, Borel parameters $M^2$, and errors in the values of the condansates. Along these lines, we calculated the coupling constant $g_{f_0\rho\gamma} = 1.97 \pm 0.57$ and $g_{a_0\rho\gamma} = 0.85 \pm 0.36$. The variation of the coupling constant $g_{f_0\rho\gamma}$ as a function of different values $M^2$ and $\theta$ are given in Fig. 4. Examination of this figure points out that the sum rule is rather stable with these reasonable variations of $M^2$. In Fig. 5 the relation between different values $M^2$ and $s_0$ is given for the variation of the coupling constant $g_{a_0\rho\gamma}$.

### IV. Conclusions

In this study we calculated coupling constants $g_{f_0\rho\gamma}$ and $g_{a_0\rho\gamma}$ in the framework of light cone sum rules in which we took account $u$- and $d$-quark contribution. Generally, even the light-cone sum rules works much better for heavy kuarks but it can also give good results for light quarks in some calculations. We applied LCSR method by using the most correct wavefunctions. In spite of lacking experimental data on $g_{f_0\rho\gamma}$, we found estimated values for the coupling constant $g_{f_0\rho\gamma}$ and $g_{a_0\rho\gamma}$ in the light-cone sum rules. The results depend on mixing angle $\theta$ and $s_0$ parameter for $g_{f_0\rho\gamma}$. When one is used reasonable data respect to analytical expressions, it is clear that one has beter agreement to experimental results. For the time being there is no experimental data on $f_0\rho\gamma$ − vertex, then our calculations behave only a theoretical suggestion. If we choose the current as $J_{f_0} = \frac{1}{2}\left(\bar{u}^b u^b + \bar{d}^b d^b\right)$ one then has to change Eq(6) so in that case we get $g_{f_0\rho\gamma} = 2.76 \pm 0.79$.

**Acknowledgments**

This work partly supported by the Research Fund of Karadeniz Technical University, under grant contract no 2002.111.001.2.

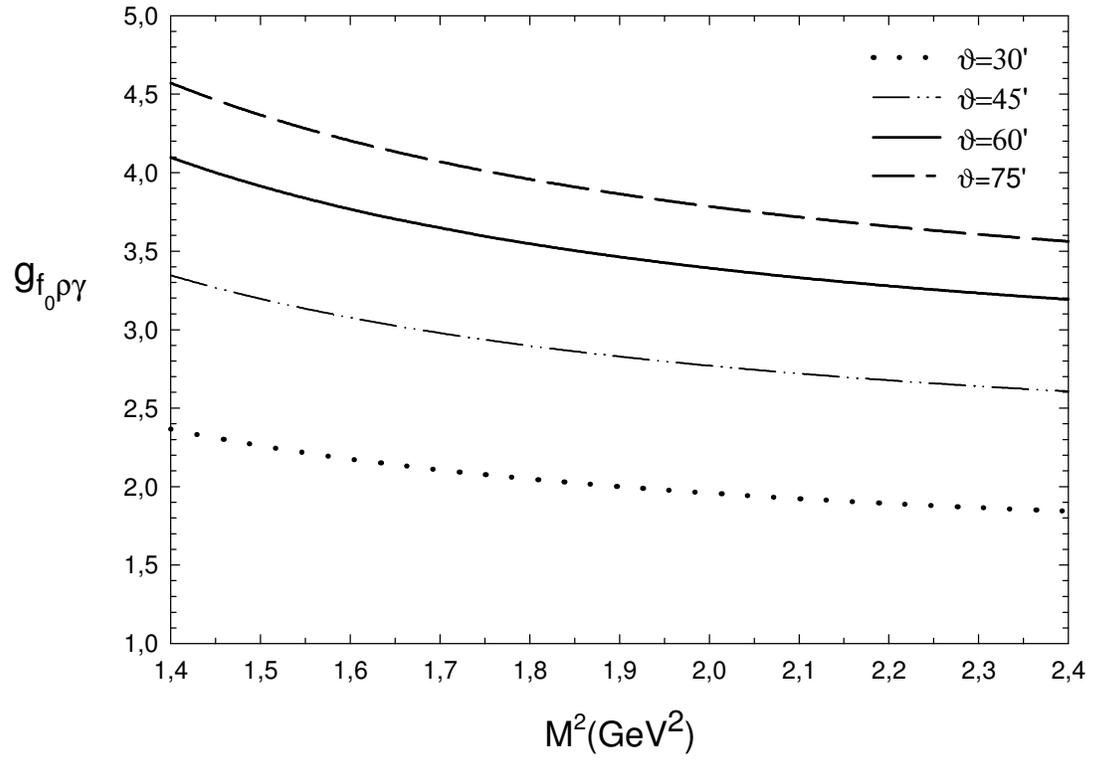

Figure 1. The variation of the coupling constant $g_{f_0\rho\gamma}$ as a function of Borel parameter $M^2$ at different $\theta$ and constant value of the continuum threshold as $s_0 = 2.0$

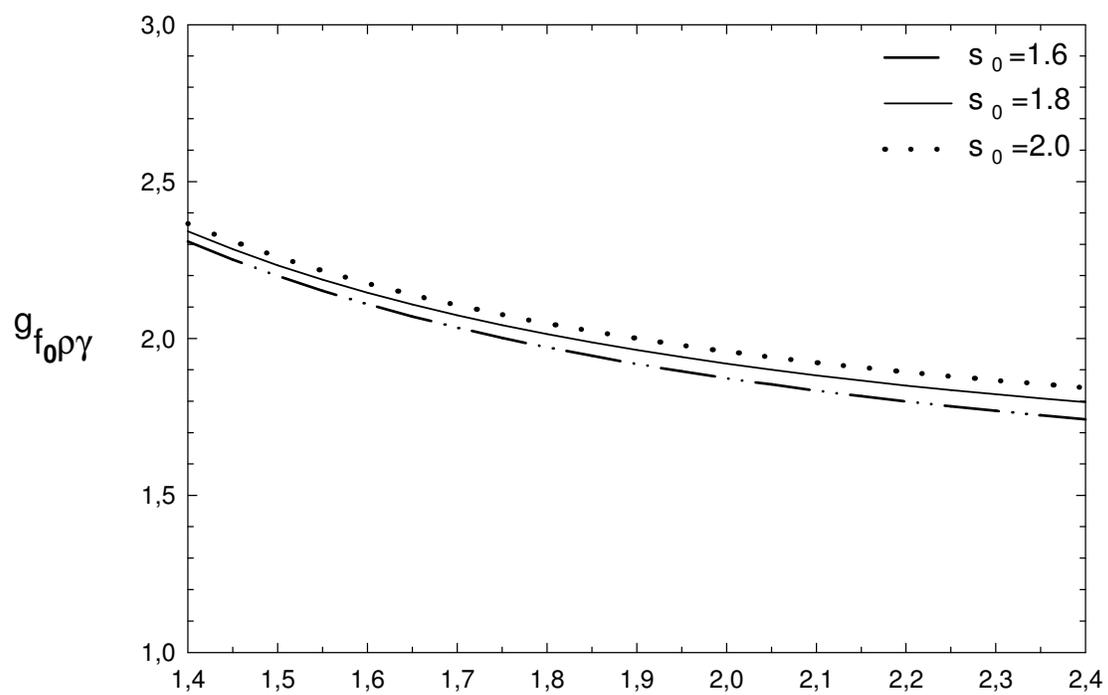

Figure 2. The dependence of the coupling constant $g_{f_0\rho\gamma}$ on parameter $M^2$ at some different values of the continuum threshold as $s_0 = 1.6$, $1.8$ and $2.0\,\text{GeV}^2$ at $\theta = 30°$.

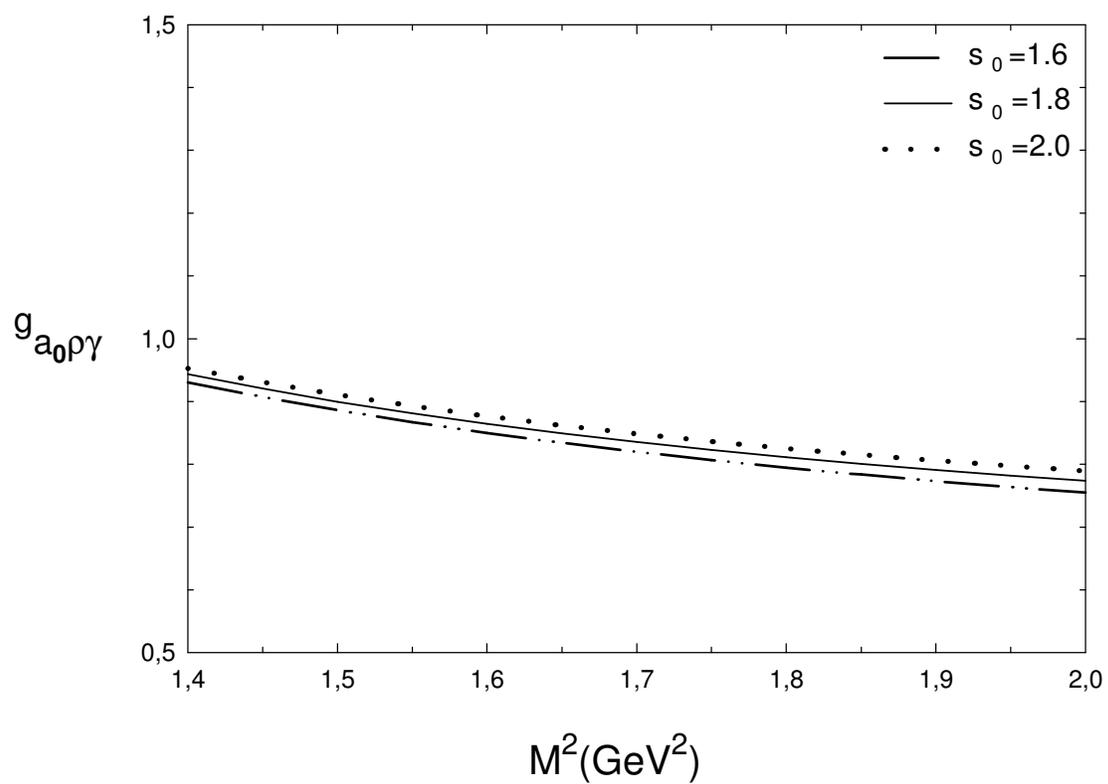

Figure 3. The dependence of the coupling constant $g_{a_0\rho\gamma}$ on parameter $M^2$ at some different values of the continuum threshold as $s_0 = 1.6$, $1.8$ and $2.0\,\text{GeV}^2$.

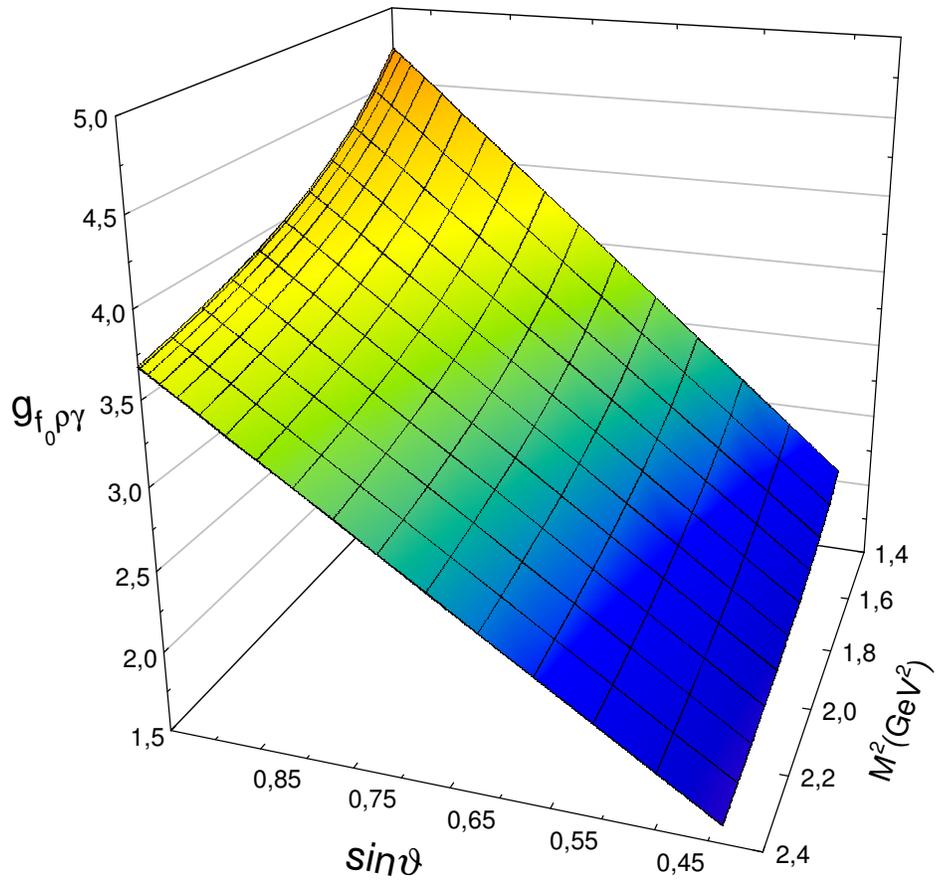

Figure 4. Coupling constant $g_{f_0\omega\gamma}$ as functions of $M^2$ and $\sin\theta$.

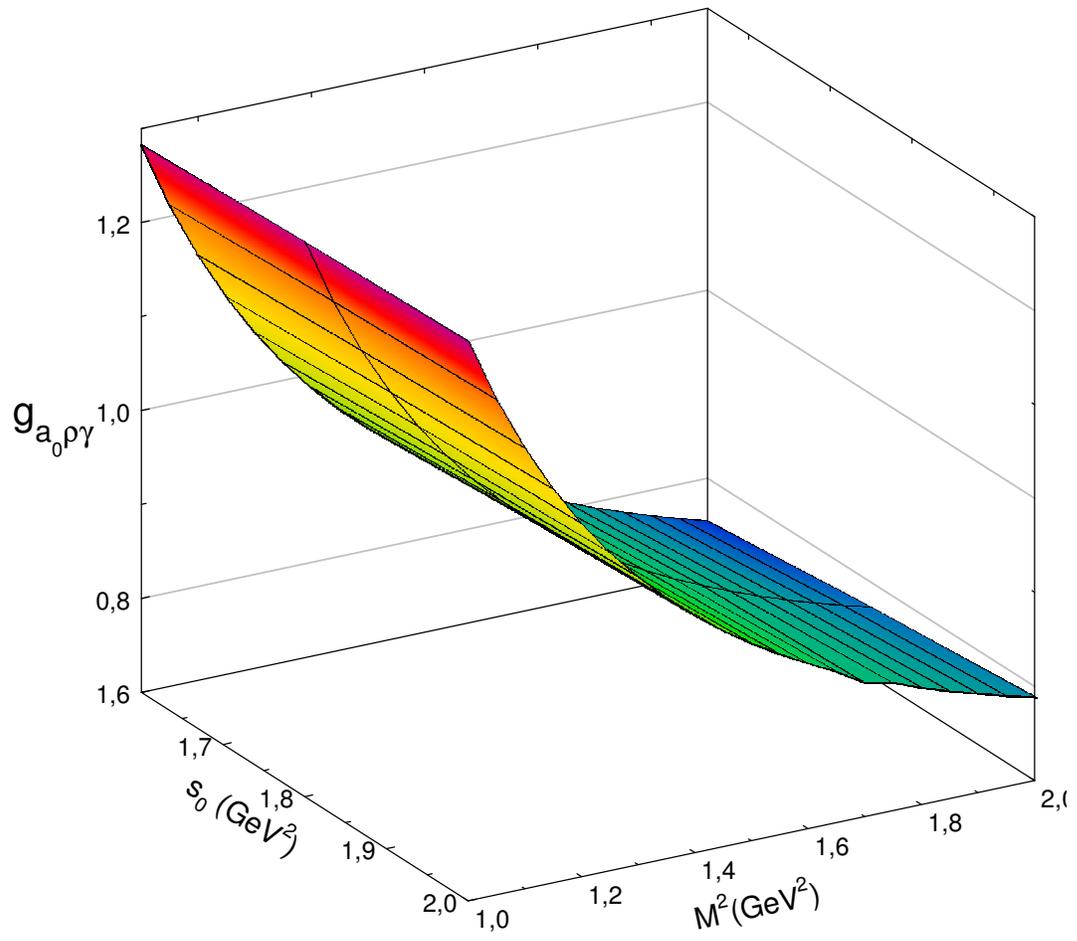

Figure 5. Coupling constant $g_{a_0\rho\gamma}$ as functions of $M^2$ and $s_0$.